# AN IR PHOTODETECTOR USING AN OPTICALLY COOLED MICROMIRROR AS A LIGHT PRESSURE SENSOR

**Gennady P. Berman, Boris M. Chernobrod**

*Theoretical Division, Los Alamos National Laboratory MS 213, Los Alamos, New Mexico 87545*

**Alan R. Bishop**

Theory, Simulation & Computation Directorate, MS B210, Los Alamos National Laboratory, Los Alamos, New Mexico 87545

**Umar Mohideen**

Department of Physics and Astronomy, University of California, Riverside, CA 92521

## Abstract

We consider mid-infrared (5 to 25 µm), optically cooled detectors based on a microcantilever sensor of the radiation pressure. A significant enhancement of sensitivity is achieved due the combination of low effective temperature (10 K), non-absorption detection, and a high quality optical microcavity. Applications to spectrometry are examined. It is shown that an optically cooled radiation pressure sensor potentially has a sensitivity an order of magnitude better than the best conventional uncooled detectors.

## I. Introduction

Recent progress in the optical cooling of microelectromechanical systems (MEMS) to millikelvin [1-3] provides a new capability for sensitivity enhancement in a variety of applications, including high-precision actuators, bolometers, magnetometers, and narrowband mechanical filters. Recently we proposed the use of a micromirror as an IR photosensor and as a THz radiation pressure sensor [4]. One of the significant advantages of radiation pressure measurements is the possibility of using a high quality microcavity, which leads to a significant sensitivity enhancement in a broad spectral region. The sensitivity limit of our proposed sensor is defined by the micromirror temperature. A lower temperature corresponds to lower mechanical noise, and higher the sensitivity of the photosensor. Conventional cooling systems are usually cost prohibitive and incompatible with many applications. Optical cooling is very desirable because it can provide compact, cost efficient, and reliable systems. In this paper we analyze the possibility of sensitivity enhancement of a previously proposed micromirror photosensor by utilizing optical cooling.

## II. Sensitivity analysis of a narrow band heterodyne detector

Figure 1 shows a possible setup for a microcantilever-based narrowband detector as a radiation pressure sensor. This optical scheme is similar to that used for efficient laser cooling of microcantilevers [1,2], which, in principle, could cool these microcantilevers to their quantum mechanical ground states. The radiation of the heterodyne laser is mixed with the signal and sent through an optical waveguide, the other end of which is polished and coated with a high reflectivity material.

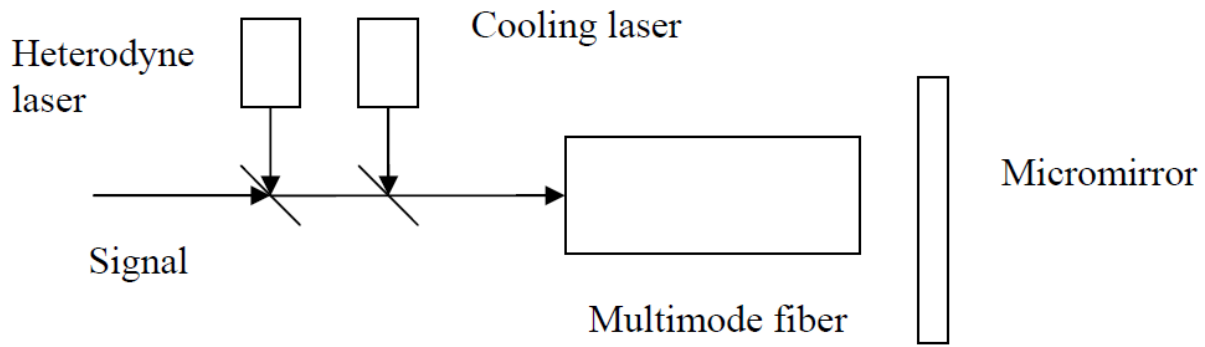

Figure 1: Schematic setup for a microcantilever-based narrowband detector.

The waveguide coating, in combination with the coated surface of the cantilever, forms a Fabry-Perot optical resonator. The second laser has a different wavelength for cooling the micromirror in the Fabry-Perot resonator.

The cantilever oscillations are measured by a Michelson interferometer. We use a model that describes the oscillations of the cantilever as a damped harmonic oscillator driven by light pressure and thermal noise. The fields in the Fabry-Perot cavity are described by the resonator equations which include damping and incident waves. The electromagnetic fields are written in the form:

$$E = \frac{1}{2}\left(E_h(t)\exp(-i\omega_h t) + E_s(t)\exp(-i\omega_s t) + E_c \exp(-i\omega_c t) + c.c.\right), \qquad (1)$$

where $E_{h,s,c}(t)$, $\omega_{h,s,c}$ are the amplitude and the frequency of the heterodyne, signal, and cooling fields, correspondingly. The slow field amplitudes inside the resonator satisfy the following equations:

$$\frac{dE_{h,s,c}}{dt} = -\frac{1}{\tau_p}\left(E_{h,s,c} - E^0_{h,s,c}(t)\right), \qquad (2)$$

where the damping time of the resonator is $2\tau_p^{-1} = (1-R)c/(2L\sqrt{R})$, $R = R_1 R_2$, $R_{1,2}$ are the reflection coefficients of the fiber end and the microcantilever surface, correspondingly, and $L$ is the average resonator length. For steady-state, the field amplitudes, $E^0_{h,s,c}$, are given by the expressions:

$$E^0_{h,s,c} = \frac{E^{ext}_{h,s,c} T_0}{1 - R \exp(\delta_{h,s,c})}, \qquad (3)$$

where $T_0$ is the transparency coefficient of the waveguide end; $E^{ext}_{h,s,c}(t)$ are the external fields launched into the resonator (below, we will consider a more general case when $E^0_{h,s,c}$ are time-dependent); $\delta_{h,s} = 2k_{h,s,c}(L+x)$ is the phase of the round-trip pass through the resonator; $x$ is the coordinate of the microcantilever; and $k_{h,s,c}$ are the wave numbers.

The motion of the microcantilever is described by the equation for a harmonic oscillator perturbed by radiation pressure, thermal noise, and cooling forces:

$$\ddot{x} + \Gamma \dot{x} + \omega_0^2 x = A|E_h|^2 + A\left[E_h^* E_s \exp(-i\omega_0 t) + c.c.\right] + \frac{F}{m} + \int_0^t \frac{dF_c(z(t'))}{dt'} h(t-t'), \qquad (4)$$

where $A = \dfrac{S}{4\pi m}$; $S$ is the area of microcantilever surface; $m$ is the microcantilever effective mass; $\Gamma$ is the damping rate; and $F$ is the thermal noise force. The integral on the right side of Eq. (4) is the light-induced force, which is assumed to be proportional to the light stored in the cavity. This force includes the radiation pressure and thermoelastic, and photothermal (bolometric) forces. $h(t-t')$ is the delayed response function. It is known that the action of the cooling force leads to a renormalization of the damping rate, $\Gamma$, frequency, $\omega_0$, and the effective temperature of the micromirror, $T_{eff}$ [1]. We take into account the fact that the temperature of the micromirror is significantly lower than the temperature of the environment. Thus, in further consideration we use Eq. (4) without the cooling force. We assume that the difference between the heterodyne frequency and the signal carrier frequency is equal to the micromirror fundamental frequency, $\omega_0 = \omega_h - \omega_s$. In Eq. (4) we omit terms corresponding to the interference between the heterodyne/signal fields from one side and the cooling field from the other side, assuming that the difference between their frequencies is far from resonant with the fundamental micromirror frequency. We also assume that the intensity noise of the heterodyne and cooling lasers are negligibly small. Thus, the first term on the right side of Eq. (4) is a constant. It is known [1] that this term leads to a steady-state shift of the cantilever amplitude to a new point of equilibrium, and it changes the frequency and damping rate of the cantilever oscillations. Below, we consider the time-dependent part of the amplitude, $x$, only. We take into account the phase fluctuations of the heterodyne laser field. (See Eq. (7).) We assume that the single mode heterodyne and signal lasers have a Lorentzian form of the spectrum (7). To obtain an analytic solution, we use the Fourier transform equations corresponding to Eqs. (2)-(4). The Fourier transform of the interference term (the second term on the right side of Eq. (4)) gives the convolution of the Fourier amplitudes of the heterodyne and signal fields:

$$i\omega E_{h,s}(\omega) = -\frac{1}{2\tau_p}\left(E_{h,s}(\omega) - E_{h,s}^0(\omega)\right), \tag{5a}$$

$$\left(\omega_0^2 - \omega^2 - i\omega\,\Gamma\right)x(\omega) =$$
$$A\int_{-\infty}^{+\infty} d\omega_1\left[E_h(\omega_1)E_s^*(\omega + \omega_0 + \omega_1) + E_h^*(\omega_1)E_s(\omega_1 + \omega_0 - \omega)\right] + F(\omega)/m, \tag{5b}$$

where the following Fourier transformations were used,

$$E_{h,s}(\omega) = \int E(t)e^{-i\omega t}dt, \quad x(\omega) = \int x(t)e^{-i\omega t}dt. \tag{6}$$

We assume that the signal is the stationary broadband field emitted by a thermal object. The $\delta$-correlated fields are given by the expression:

$$\frac{cS}{8\pi}\left\langle E_{h,s}^{ext}(\omega)E_{h,s}^{ext*}(\omega')\right\rangle = P_{h,s}\frac{\Gamma_{h,s}}{\pi(\omega^2+\Gamma_{h,s}^2)}\delta(\omega-\omega'), \tag{7}$$

where $P_{h,s}$ is the signal power incident on the resonator. For the spectral components of the thermal noise force, we have:

$$\left\langle F(\omega)F^*(\omega')\right\rangle = \delta(\omega-\omega')\frac{k_B T_{eff} K\Gamma}{\omega_0^2 \pi}, \tag{8}$$

where $K = m\omega_0^2$ is the spring constant of a microcantilever; $k_B$ is the Boltzmann constant; and $T_{eff}$ is the effective mirror temperature.

Below, we calculate the minimal measurable spectral irradiance (MMSI) using the equality:

$$\left\langle x_s^2(t)\right\rangle = \left\langle x_T^2(t)\right\rangle, \tag{9}$$

where $\left\langle x_s^2(t)\right\rangle$ is the mean square amplitude, averaged over the field fluctuations, of oscillations induced by the radiation pressure; and $\left\langle x_T^2(t)\right\rangle = k_B T_{eff}/K$ is the thermal noise mean square amplitude. Note that the system of equations for the harmonic oscillator and the field in an optical cavity (Eqs. 5a,b) is nonlinear due to the nonlinear dependence of the field amplitude in the resonator on the microcantilever coordinate given by Eq. (3). We consider a linear approximation to the solution of Eqs. (5a,b). We assume that, for realistic values of parameters, the oscillation amplitude of the cantilever is much smaller than the region of dispersion of the optical resonator. In this paper we are not interested in the nonlinear regime of the cantilever vibrations (when the amplitude of the cantilever vibrations is larger than the length of dispersion of the optical resonator). The equilibrium position of the microcantilever can be chosen to provide the maximal field enhancement inside the resonator. In this case Eq. (3) gives

$$E^0_{h,s}(\omega) = \frac{E^{ext}_{h,s}(\omega)T_0}{1-R}. \qquad (10)$$

We assume that the line widths of the heterodyne and signal sources, $\Gamma_{h,s}$, the micromirror resonance frequency, $\omega_0$, and the damping rate, $\Gamma$, are much less than the bandwidth of the optical resonator, $\tau_p^{-1}$: $\Gamma_{h,s}, \omega_0, \Gamma \ll \tau_p^{-1}$. The frequency deviation of the signal spectral components from the heterodyne carrier frequency is of the order of $\omega_0$. (See Eq. (12) below.) These frequency differences are negligibly small compared with the resonator line width. Thus, Eq. (10) can be satisfied simultaneously for the heterodyne field and for the signal field.

To calculate the mean square amplitude, $\langle x_s^2(t) \rangle$, we use the solution of Eqs. (5a,b):

$$x_s(t) = A\int_{-\infty}^{\infty} d\omega_1 \int_{-\infty}^{\infty} d\omega \frac{\exp(-i\omega_1 t)(E_h(\omega)E_s^*(\omega+\omega_0+\omega_1) + E_h^*(\omega)E_s(\omega+\omega_0-\omega_1))}{\omega_0^2 - \omega_1^2 - i\omega_1\Gamma}. \qquad (11)$$

To calculate the average of amplitude squared, $\langle x_s^2(t) \rangle$, we have to take into account the fact that the spectral components of the fields, $E_{h,s}(\omega)$, are δ-correlated. Performing the two integrations with δ-functions, and taking into account Eqs. (7,8,10,11), we obtain the following expression for the mean square amplitude:

$$\langle x_s^2(t) \rangle = B \int_{-\infty}^{\infty} d\omega_1 \int_{-\infty}^{\infty} d\omega \frac{P_s(\omega + \omega_0 - \omega_1)}{(\omega^2 + \Gamma_h^2)(\omega^2 + \tau_p^{-2})[(\omega + \omega_0 - \omega_1)^2 + \tau_p^{-2}][(\omega_1^2 - \omega_0^2)^2 + \omega_1^2 \Gamma^2]}, \quad (12)$$

where,

$$B = \frac{T_0^4 P_h \Gamma_h}{4\pi m^2 c^2 (1-R)^4 \tau_p^4}, \quad P_s(\omega + \omega_0 - \omega_1) = P_s \frac{\Gamma_s}{\pi[(\omega + \omega_0 - \omega)^2 + \Gamma_s^2]}.$$

The integrals in Eq. (12) are calculated assuming that $\Gamma_h \ll \tau_p^{-1}$. Performing these straightforward calculations, we obtain

$$\langle x_s^2(t) \rangle = \frac{4 P_h P_s \omega_0^2 T_0^4}{c^2 K^2 \Gamma (\Gamma_h + \Gamma_s)(1-R)^4}. \quad (13)$$

Combining Eqs. (9) and (13) we obtain for the MMSI

$$MMSI = \frac{k_B T_{eff} \rho d c^2 \Gamma \left(1 + \frac{\Gamma_h}{\Gamma_s}\right)(1-R)^4}{2 P_h T_0^4}, \quad (14)$$

where $\rho$ is the density of the cantilever material and $d$ is the thickness of the micromirror. According to Eq. (14) the most critical parameter is the coefficient of reflection, $R$. Micromirrors with reflectance better than 0.99 for infrared radiation are commercially available. A micromirror with reflectance 0.9999 was demonstrated recently in laser cooling experiments [3]. In our estimate we use the value, $R = 0.996$. The microcavity with flat mirrors, such as used in [1], has diffraction losses (leak losses) which can exceed the transmission losses for a very high reflectivity. The efficient solution of this problem was realized in [5], where the mirror of the Fabry-Perot microresonator consists of movable and unmovable parts. Because the oscillation amplitude of the movable part of the mirror is negligibly small, it does not affect the resonator losses. At the same time, the total area of the mirror is large enough to avoid diffraction losses. To obtain higher sensitivity we require a thinner mirror. However, the mirror technology limits the thickness depending on the size of the mirror. Thus, for a movable mirror with smaller area, a smaller thickness can be manufactured. For example, using the movable mirror size of the order of, $100 \times 10$ μm$^2$, we can assume that the thickness is of the order of 60 nm. Note that a microcavity with curved mirrors in stable paraxial geometry, such as used in [2], also overcomes the problem of diffraction losses. For the effective temperature we assume 10 K. This temperature can be achieved by laser cooling starting from room temperature [2]. Typical values of parameters are: $\rho = 2.33 \times 10^3 kg/m^3$, $d = 60\ nm$, $\Gamma = 2\pi \times 13\ s^{-1}$, $\Gamma_h = \Gamma_s$, $T = 10\ K$, $R = 0.996$, $T_0^2 = (1-R)^2$, and $P_h = 50$ mW. For these values of the parameters, Eq. (14) gives $MMSI = 4.5 \times 10^{-11}\ W/m^2 Hz$.

For comparison of the sensitivity of the proposed spectrometer with a standard IR-spectrometer with uncooled detector, consider a numerical example. We assume that a conventional IR spectrometer is equipped with a photosensor array, in which an individual pixel is a microbolometer. We assume also that the conventional spectrometer operates in the spectral interval, 8- 14 $\mu$m, and has the same spectral resolution as the proposed spectrometer. In our case, the spectral resolution is defined by the laser line width, which we choose equal to $\Delta \nu = 30$ MHz. For a conventional spectrometer, this high spectral resolution can be achieved in combination with a Fabry-Perot etalon. The theoretical limit for noise equivalent thermal difference (NETD) of a microbolometer with pixel size of $30 \times 30$ $\mu m^2$ is 5 mK. (The experimental sensitivity is 20-50 mK.) Our estimate uses NETD = 5 mK. The minimal measurable spectral irradiance is $MMSI = NETD \times (dP/dT)_{\lambda_1 - \lambda_2} \times (\Delta \nu)^{-1}$, where $(dP/dT)_{\lambda_1 - \lambda_2}$ is the slope of the blackbody radiation within the spectral band 8-14 $\mu m$: $(dP/dT)_{\lambda_1 - \lambda_2} = 2.62$ $Wm^{-2} K^{-1}$. For $\Delta \nu = 30$ MHz, NETD = 5 mK, we get $MMSI = 4.4 \times 10^{-10}$ W/(m$^2$ Hz). Thus, the microcantilever spectrometer has a sensitivity that is an order of magnitude better than the best conventional uncooled detectors.

### III. Conclusion

We have described an infrared detector based on an optically cooled micromirror as a radiation pressure sensor. Our theoretical analysis shows that this radiation pressure sensor potentially has sensitivity better than the best conventional uncooled detectors.

AC52-06NA25396. G.P.B and U.M were also supported by a UC Lab Fees Research Program grant.